\documentclass{egpubl}
\usepackage{eurovis2026}
\usepackage{tikz}
\usepackage[most]{tcolorbox}
\usepackage{xcolor}
\newcommand{\highlightcolor}{black}
\newcommand{\colortext}[1]{\textcolor{\highlightcolor}{#1}}
\definecolor{dccolor}{RGB}{65, 105, 225}  

\newtcolorbox{dc}[1]{
  enhanced,
  borderline west={2.5pt}{0pt}{dccolor},
  boxrule=0pt,
  frame hidden,
  colback=dccolor!3,
  sharp corners,
  left=8pt,
  right=0pt,
  top=2pt,
  bottom=2pt,
  fonttitle=\bfseries\small,
  coltitle=dccolor,
  title={#1},
  attach title to upper={\quad}, 
  before skip=8pt,
  after skip=8pt
}

\SpecialIssuePaper         

\CGFccby

\usepackage[T1]{fontenc}
\usepackage{dfadobe}  

\usepackage{cite}  
\BibtexOrBiblatex
\electronicVersion
\PrintedOrElectronic

\usepackage{egweblnk}
\title
      {Input Visualizations to Track Health Data by Older Adults with Multiple Chronic Conditions}

\author[Ramesh et al.]
{\parbox{\textwidth}{\centering 
  Shri Harini Ramesh$^{1}$\orcid{0009-0002-4245-7693},
  Foroozan Daneshzand$^{1,2}$\orcid{0009-0001-8395-8948},
  Matteo Sotelo$^{3}$,
  Mahsa Sinaei$^{3}$\orcid{0009-0004-4424-9016},
  and Fateme Rajabiyazdi$^{1}$\orcid{0000-0002-8710-865X}
  }
\\
{\parbox{\textwidth}{\centering 
  $^1$Department of Computer Science, University of Calgary, Canada\\
  $^2$School of Computing Science, Simon Fraser University, Burnaby, British Columbia, Canada\\
  $^3$Department of Systems and Computer Engineering, Carleton University, Canada
  }
}
}
\begin{document}
\teaser{
  \includegraphics[width=0.9\linewidth, alt={Physical input visualizations of health data created by our nine study participants}]{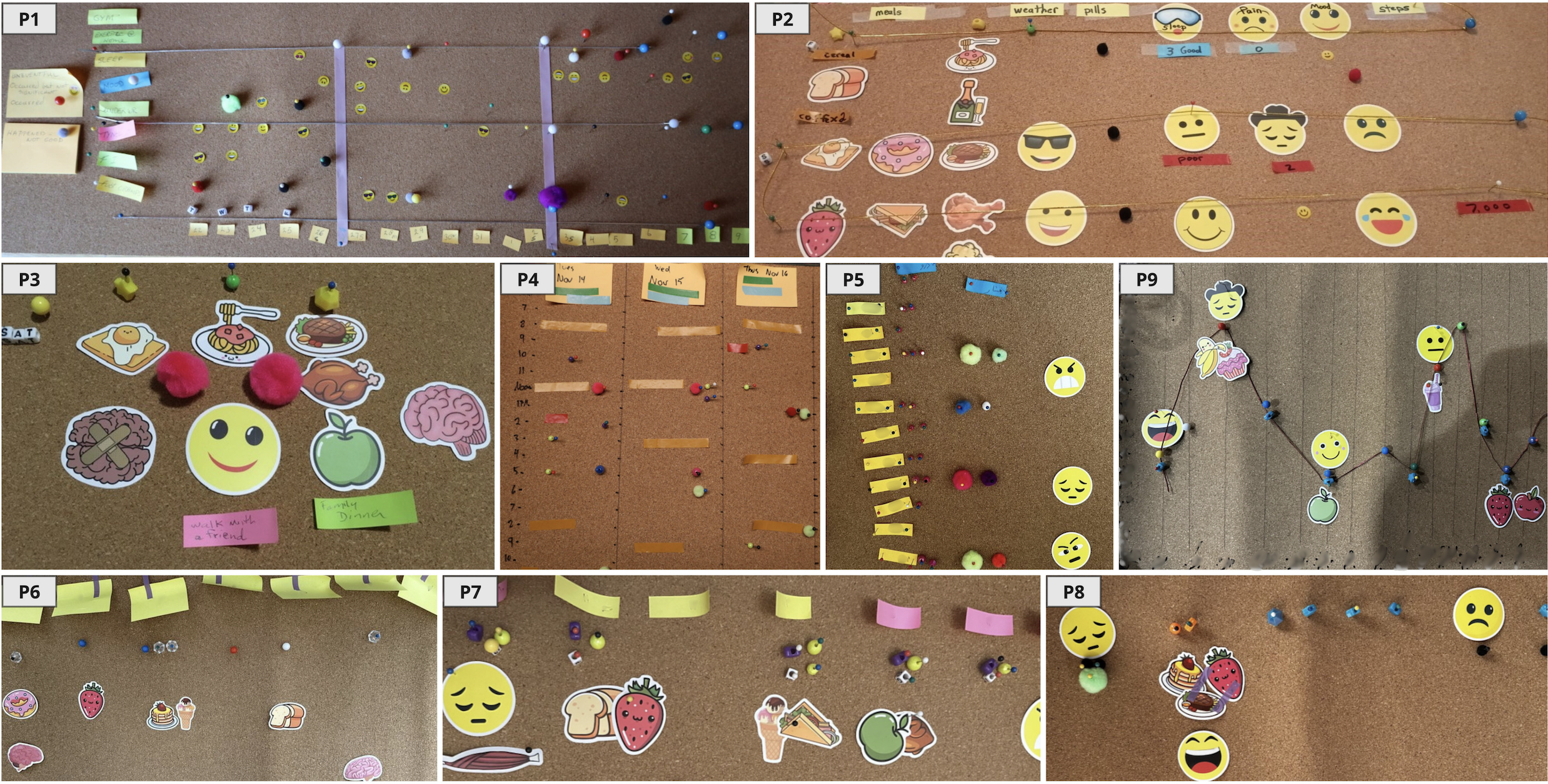}
    \centering
    \caption{%
  	Physical input visualizations of health data created by our nine study participants (older adults living with MCC).%
  }
  \label{fig:teaser}
}

\maketitle
\begin{abstract}
Older adults living with multiple chronic conditions (MCC) can considerably benefit from collecting and reflecting on their health data. 
\colortext{Many older adults collect their health data using various approaches, such as digital tools or handwritten notebooks. 
However, in these approaches, the act of collecting data does not itself yield insights; sensemaking and reflection happen only if individuals later review their accumulated records. 
The daily process of data collection thus offers limited opportunity for individuals to actively engage with their data or find the process personally meaningful and enjoyable.}
Personal data input visualizations using physical tokens offer a promising solution that can help individuals recognize evolving patterns while collecting data and discover meaningful insights in a more serendipitous and engaging manner.
Yet, there is a limited understanding of whether and how older adults living with MCC might adopt physical input visualizations to collect data and reflect on their health, and how the tangible, expressive, and personalizable nature of this process supports their sensemaking and reflection.
In this paper, we present the results of our interview and diary studies in which older adults living with MCC inputted health data using physical tokens over two weeks. 
Our findings highlight the diverse and unique needs of older adults for tracking personal health data, illustrating how they adapt strategies and personalize physical input visualizations to align with their individual needs. We demonstrate how older adults integrated input visualizations into daily routines and leveraged tangible markers to reflect on patterns and behaviors, while enjoying the process of tracking and focusing on personal expression and meaningful reflection.
Finally, we provide design considerations for supporting older adults with MCC when inputting health data through physical tokens. All data and supplemental materials are available at~\url{https://osf.io/7ak9x/overview?view_only=63ce1356b4b94a1abea6d9247c8a2f26}.


\end{abstract}  
\section{Introduction}
The use of patient-generated health data (PGHD) plays a pivotal role in self-monitoring and managing multiple chronic conditions (MCC)~\cite{Ploeg2019}. 
\colortext{Making sense of the patterns that emerge from PGHD and reflecting on daily health behaviors can empower individuals to better understand their health and make informed decisions about their self-care and management~\cite{Lim2019}. 
Many older adults living with MCC collect their health data using various approaches such as digital tools or handwritten notebooks~\cite{Oh2022}.
However, older adults often encounter barriers with these approaches, such as difficulty personalizing digital tools for their unique and evolving health conditions~\cite{Zaman2022} and the lack of visual support in handwritten records to discern patterns from numerical data~\cite{GerentoVis2024Euro}. 
Moreover, in all these approaches, the act of collecting data itself does not yield insights; that is, sensemaking happens only if individuals later review their accumulated records of health data. 
Thus, the daily process of data collection offers limited opportunity for individuals to actively engage with their data, make sense of emerging patterns, or find the process personally meaningful and enjoyable.
There is a need for approaches that make the process of collecting health data itself engaging, enjoyable, and insightful, where individuals can see patterns emerge and reflect on their health as they collect.
}

\colortext{\textit{Input visualization}, the act of collecting data with visual representations, supports individuals in seeing patterns and gaining insights as they emerge during data collection, making the tracking process engaging and meaningful~\cite{Bressa2024Input}.}
Using physical tokens for input visualization can further support older adults by mitigating technological barriers and providing an intuitive, personalizable way to record, represent, and reflect on data~\cite{dragicevic2021}. 
Individuals who collected personal data through physical input visualizations reported noticing changes in their behavioral patterns, making serendipitous discoveries, enjoying artistic expression, and reflecting on their behaviors~\cite{Abtahi2020}, \colortext{suggesting that physical input visualizations can foster active engagement, enjoyment, and a meaningful data collection experience that current approaches lack.} 
However, we lack an understanding of whether and how older adults living with MCC could use physical input visualizations to track and meaningfully reflect on their health data, \colortext{and what role the tangible, expressive, and personalizable nature of this process plays in supporting sensemaking and reflection.}

To fill this gap, we conducted a two-part study with 9 older adults (> 65 years old) living with three to six chronic conditions (MCC). The study entails a preliminary semi-structured interview to gain an understanding of the challenges they face in collecting and presenting their health data.  
In addition, we conducted a two-week diary study in which patients tracked their health data by inputting with visual representations using physical tokens, along with an exit semi-structured interview to reflect on their experiences.
Our findings underscore how older adults living with MCC personalized their physical input visualizations to collect health data in expressive, personally relevant ways that encouraged active engagement, supported pattern recognition, and prompted reflection. 
Patients selected tokens that were personally meaningful, and wanted to update their data on the go in places where they spent time. 
They positioned their visualizations in common living areas, dining rooms, and kitchens, to make tracking an engaging, mindful, and sustainable part of their daily routines. 
We have three contributions: 
\begin{enumerate}
    \item We highlight the diverse and unique needs of older adults living with MCC for tracking personal health data, illustrating how they adapt strategies and personalize physical input visualizations to align with their individual needs.
    \item \colortext{We demonstrate that the tangible, expressive, and personalizable nature of physical input visualizations} facilitates sensemaking and reflection by enabling older adults to recognize patterns, discern relationships in their health data, and reflect on their behaviors, while also serving as tangible reminders for self-care.
    \item We propose design considerations to support older adults with MCC in the process of 
    inputting health data with visualizations.
\end{enumerate}




\section{Related Work}
In this section, we first outline how older adults with MCC currently self-monitor and manage their health. We then explored tools designed for inputting data using physical tokens and examined tools developed specifically for self-reflection and self-expression.

\subsection{Self-monitoring and managing health}

Self-monitoring and management of health conditions involves consistently tracking and interpreting personal health data~\cite{Oh2022}, which can increase engagement in one's care, promote healthy behaviors, and support more informed decisions about self-management~\cite{Ploeg2019}. 
These benefits are especially important for older adults, many of whom live with multiple chronic conditions such as hypertension, diabetes, cardiovascular disease, arthritis, or chronic pain~\cite{Lim2019}. 
With advances in technology, some older adults use digital tools, such as smartphone apps and wearable devices, to track personal health data~\cite{Chandrasekaran2020}.
\colortext{However, older adults often face challenges when using these technologies, including age-related changes in vision and motor control that complicate screen-based interaction, high cognitive load from complex interfaces, and limited digital literacy~\cite{Zaman2022}.}
Additionally, most digital health tools offer limited personalizability, constraining older adults managing MCC from tracking their choice of data in the way they need~\cite{Lim2019}. 
\colortext{Many older adults also use handwritten notebooks, which offer familiarity but lack visual support to help discern patterns from numerical data~\cite{GerentoVis2024Euro}.
Moreover, in all these approaches, the act of collecting data remains separate from seeing patterns or understanding the data,} making meticulous tracking feel burdensome, particularly when the effort invested does not yield meaningful insights~\cite{Ancker2015}. Thus, there is a need to support older adults in collecting and interpreting health data in ways that feel intuitive and engaging rather than tedious or cognitively demanding. 
\colortext{In this paper, we explore whether physical input visualizations, where older adults use physical tokens to visually represent their health data, could offer an engaging and meaningful way for older adults with MCC to collect, make sense of, and reflect on their data for self-monitoring and managing their conditions.}

\subsection{Tools for input visualizations using physical tokens}
Physical input visualizations are visual representations designed to collect or modify new data, rather than encode existing datasets, using physical tokens~\cite{Bressa2024Input}. Inputting data through physical tokens can support cognition, understanding, communication, learning, problem-solving, and decision-making~\cite{Jansen2015}. 
Research suggests that constructing visualizations using physical tokens can empower individuals, especially non-experts, to collect data and make visualizations that are simple, expressive, flexible, and accessible to everyone~\cite{huron2022making}.
Previous studies introduced prototyping tools and toolkits that help individuals input data with visual representations using physical tokens across different contexts.
\colortext{Previous studies have introduced tools and toolkits that help individuals input data with visual representations using physical tokens across diverse contexts, including dietary choices and climate impact (Edo~\cite{Edo2023}), collaborative project documentation in makerspaces (Cairn~\cite{Cairn2017}), home sensor data (SensorBricks~\cite{Brombacher2024}), and personal self-tracking data (DataChest~\cite{Wijers2024}). }
While this line of research demonstrates how physical tokens can be used to input data across diverse contexts, little is known about the benefits of inputting data using physical tokens for older adult patients. 
\colortext{In this research, we explore the benefits of physical input visualizations for older adults with MCC in making sense of their health data as they collect it.}

\subsection{Physical input visualizations for self-reflection and self-expression}

Inputting data with physical visualizations can encourage and enable people to self-collect data and self-reflect about aspects of their personal lives~\cite{huron2022making}. 
Thudt et al.~\cite{Thudt2018} found that using physical materials to input personal data through the hands-on approach fostered deep reflection, captured subjective aspects, and supported sharing and planning.
Panagiotidou et al.~\cite{Panagiotidou2020} showed through Data Badges that physical data input facilitated freedom of creation and diverse forms of data expression. 
Wannamaker et al.~\cite{Wannamaker2021} demonstrated through I/O Bits that tangible materials help users focus on tracking tasks and create meaningful opportunities for reflection during the activity itself. 
Bentvelzen et al.~\cite{Bentvelzen2023} showed that creating tangible representations of health data using LEGO® bricks can provide engaging experiences and promote reflection. 
\colortext{These studies show that the tangible, hands-on process of constructing visual representations of data, selecting tokens, assigning personal meanings, and physically arranging tokens, fosters reflection, personal expression, and meaningful engagement in ways that go beyond simply recording data. 
However, little is known about how older adults, who have diverse and evolving needs, take advantage of these tangible, expressive, and personalizable qualities of physical input visualizations for self-reflection and self-expression, which this study investigates.}
\section{Methods}

Our goal was to gain a deep understanding of whether and how older adults living with MCC could use physical input visualizations to track and promote reflection on their health.
To achieve this, we conducted a two-part study: 1) an introductory semi-structured interview and 2) a two-week diary study along with an exit semi-structured interview. \colortext{This study was approved by the [university's] research ethics board. All participants provided written informed consent prior to the introductory interview.}

\subsection{Introductory interview}
We conducted one-hour, semi-structured interviews using open-ended questions (See Appendix A). Initially, we gathered demographic information from each participant, including age, gender, educational background, and chronic conditions. 
We then asked how they collect, manage, and interpret their health data at home. We also inquired about the challenges they face in this process.

\subsection{Diary study}

\begin{figure}
    \centering
    \includegraphics[width=0.9\linewidth]{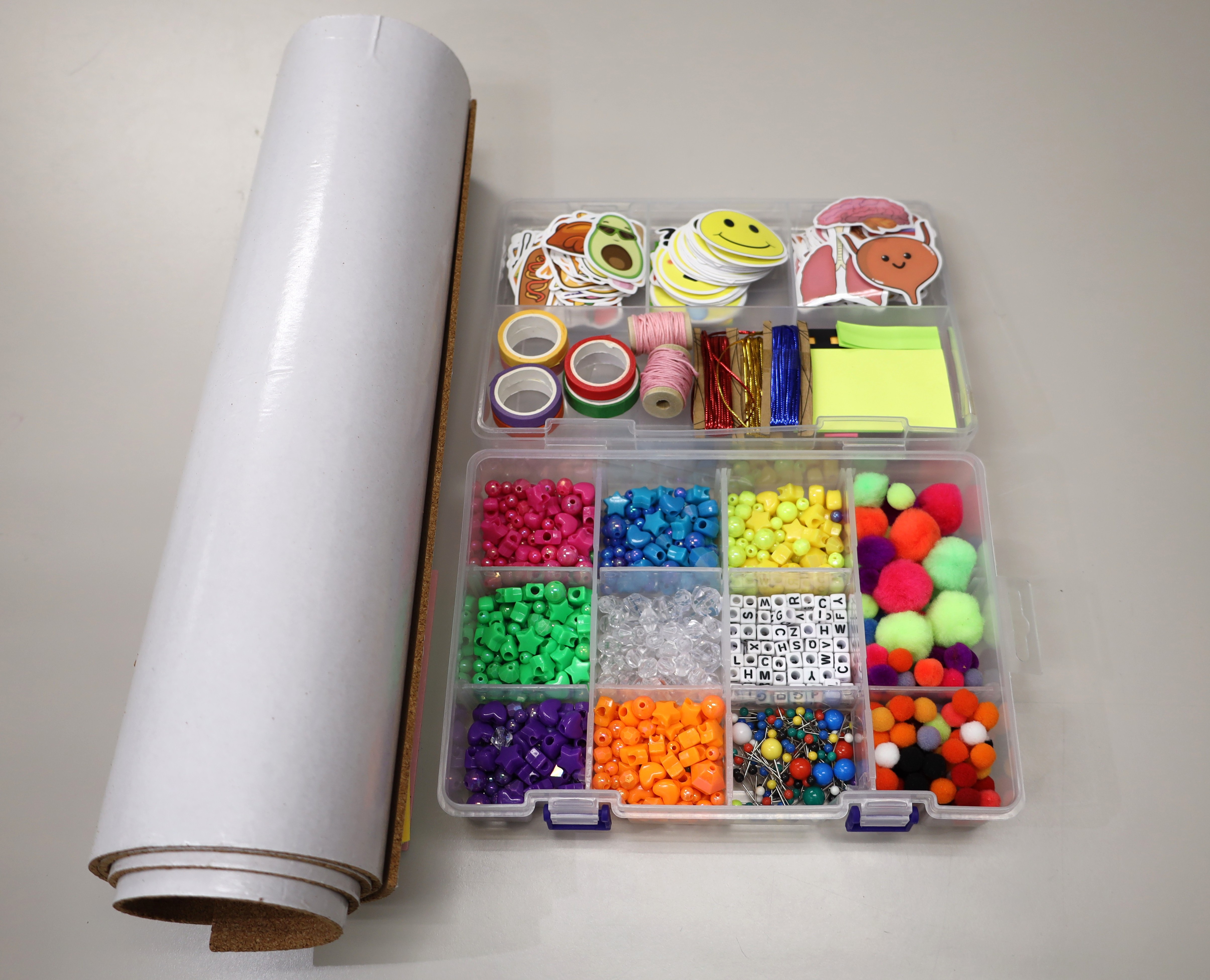}
    \caption{Toolkit for physical input visualization of health data.}
    \label{fig:physicalviskit}
\end{figure}

\textbf{Toolkit for physical input visualization of health data: }
 We developed and iteratively refined a toolkit to input health data through pilot testing with members of our research group, a group of human–computer interaction (HCI) and visualization researchers. During these pilot sessions, participants used different materials to represent sample health data, and their feedback informed successive refinements of the toolkit. Through this iterative process, we experimented with a range of materials for both the physical medium for inputting data with visual representations (e.g., foam board, cardboard, cork board) and tokens (e.g., beads, pins, stickers, and threads), drawing inspiration from prior input visualization toolkits~\cite{Wijers2024} \colortext{and literature on age-related changes in dexterity and vision~\cite{ Chandrasekaran2020} to ensure suitability for older adults. 
Finally, we selected those that were lightweight, portable, and easy to grasp and distinguish} while offering variety in size, shape, and color to support diverse forms of data representation. 
We then provided the finalized toolkit to participants for inputting data using tokens (see \textbf{Figure~\ref{fig:physicalviskit}}). This kit included a cork board (47x16 inches), beads (7 colors, 5 shapes), pins (6 colors, 2 sizes), puffy balls (11 colors, 4 sizes), tape (6 colors), thread (4 colors), sticky notes (5 colors), labels (5 colors), stickers (3 types: emojis, food, body parts), and alphabet square beads (26 letters). 

\textbf{Tutorial and cheat sheet:} We first conducted a 30-minute tutorial to explain to participants how to track their health data using our physical tokens. Subsequently, participants were provided with a fictional dataset that included step counts, weather, and mood for 5 days of the week and the toolkit. They were then asked to input the given data using the tokens and received feedback from the researchers. 
At the end of the tutorial, participants received a cheat sheet (See Appendix B) summarizing the tutorial so they could refer to it during the study period.

\textbf{Two-week diary:} Once the tutorial was completed, participants were asked to input their chosen health data using tokens for 2 weeks.
We adopted the diary study method because it facilitates a rich understanding of how participants intend to share and illustrate their data while also minimizing the effects of observation~\cite{Carter2005Diary}.
\colortext{The two-week diary period was chosen to provide sufficient time for participants to move beyond initial experimentation, and develop their own tracking strategies, while remaining feasible for older adults managing multiple chronic conditions~\cite{janssens2018qualitative}. Prior research has shown that within two weeks, individuals tracking health symptoms can begin to identify triggers, recognize patterns, and make connections between their behaviors and health outcomes~\cite{TURNERMCGRIEVY20191516}.}
During the study, participants were encouraged to document their visualizations via photos and notes. We also sent two email check-ins to inquire about any challenges they encountered and invited them to share photos, notes, and thoughts.


\textbf{Exit interview:} After the two-week diary, we visited participants at their residences for an in-person exit interview.  We conducted one-hour semi-structured interviews (See Appendix C), asking participants to describe how they designed and constructed their visualization, the rationale behind choosing particular data variables and representations (tokens), and the challenges they encountered throughout the process. 

\subsection{Recruitment and participants} 
We aimed to recruit a diverse group of older adults with various demographics, health conditions, and residence arrangements; thus, we used several recruitment methods to broadly distribute our call for participation.
We shared recruitment posters on social media, in the e-newsletters (sent to senior groups in the country), and on physical poster boards of several retirement homes, and employed a snowball sampling approach~\cite{palinkas2015purposeful}. Participants were required to be over 65 years old, diagnosed with at least two chronic conditions, and fluent in English or the local language. Individuals with cognitive disabilities were excluded, as such conditions could affect their ability to provide informed consent and complete the study.
We recruited nine older adult individuals living with MCC who participated in both parts of our study 
(4 Women, 5 Men; mean age \(69 \pm 3.1\) years). Most patients were
living with more than 3 chronic conditions; detailed demographics and health conditions are summarized in \textbf{Table~\ref{tab:participant-demographics}}.

\begin{table*}[ht]
\centering
\caption{Participant demographics and health conditions.}
\label{tab:participant-demographics}
\small
\renewcommand{\arraystretch}{1.2}
\begin{tabular}{|l|l|l|l|l|p{9.6cm}|}
\hline
\textbf{ID} & \textbf{Age} & \textbf{Gender} & \textbf{Ethnicity} & \textbf{Education} & \textbf{Health Conditions} \\ \hline
P1 & 69 & Man & White & College/Associate  & Hypertension, Depression, Gastroesophageal Reflux Disease, Hypercholesterolemia, Hieroglif Spastic Paraparesis \\ \hline
P2 & 68 & Woman & White & Bachelor & Chronic Lymphocytic Leukemia (Cancer), Chronic Respiratory Disease, Arthritis, Bronchiectasis, Primary Immunodeficiency, Immunoglobulin G \\ \hline
P3 & 70 & Woman & White & Master & Hypothyroidism, Familial Hypercholesterolemia, Sleep Apnea, Osteoporosis, Chronic Respiratory Disease (Asthma)\\ \hline
P4 & 65 & Man & White & Master & Type 1 Diabetes, Chronic Respiratory Disease (Asthma), Sleep Apnea, Obesity \\ \hline
P5 & 68 & Woman & White & High School  & Hypertension, Type 2 Diabetes, Chronic Back Pain, Hypercholesterolemia \\ \hline
P6 & 75 & Man & Middle Eastern & High School  & Diabetes, Cardiovascular Disease (Stroke), Early Stage of Dementia \\ \hline
P7 & 65 & Man & Middle Eastern & College/Associate  & Hypertension, Hypercholesterolemia, Type 2 Diabetes, Cardiovascular Disease \\ \hline
P8 & 69 & Woman & Middle Eastern & Middle School  & Depression, Arthritis, Osteoporosis \\ \hline
P9 & 71 & Man & Middle Eastern & PhD  & Type 2 Diabetes, Obesity, Chronic Back Pain \\ \hline
\end{tabular}
\end{table*}

\subsection{Data collection and analysis}
\textbf{Introductory interview: }We audio-recorded the introductory interviews. We transcribed the audio and translated it into English if conducted in the participant's local language. To analyze this data, we used deductive coding~\cite{merriam2002introduction}. We established three sub-themes based on the introductory interview questions. Two authors independently reviewed and coded the transcripts based on the pre-defined sub-themes using Microsoft Excel\texttrademark{}. \colortext{ Coders then compared their codes, discussed discrepancies with the principal investigator, and refined the codes.}
(See Appendix D). 

\textbf{Diary study: }Throughout the diary study, we collected participants' diary entries by picture updates via email. 
We audio- and video-recorded exit interviews and transcribed them to capture both verbal explanations and physical gestures, such as pointing to specific areas on the board or referencing particular tokens. After the exit interview, we also photographed the final physical input visualizations in the participants' residences (See \textbf{Figure~\ref{fig:teaser}}).
This triangulation of data~\cite{Carter2014TheUO}, linking participants’ verbal explanations (from audio transcripts) with their gestures (from video) and the corresponding pictures (from physical boards), enabled us to identify which parts of the board they referenced, the tokens they used to represent their data, and their rationale for constructing physical representations. 
To analyze the data, we conducted an iterative inductive thematic analysis approach \cite{thematicanalysis}. \colortext{Each transcript was independently analyzed by three authors to identify emerging themes and patterns within the data. Subsequently, the coders met to discuss the identified codes, examining their similarities and differences. Any disagreements or uncertainties in coding were discussed and resolved during the coding process on a case-by-case basis. }
Through this process, we gathered \textcolor{black}{65} codes. 
We then applied axial coding to organize and connect the initial codes, establishing relationships and patterns across the data~\cite{corbin2014basics}. Finally, we grouped these codes into eleven sub-themes and three overarching themes, \colortext{forming the final codebook} (See Appendix E).

\section{Introductory Interview Results}

Our analysis of the introductory interviews revealed how patients with MCC collect and reflect on their data and the challenges they encounter.
Some patients in our study collected health data, such as blood pressure, blood sugar, and medication dosages, using pen-and-paper or digital tools. However, many faced challenges in interpreting these data and reflecting on their behaviors. 
Several patients did not collect any data beyond collecting their clinician-generated test and laboratory results. 
Four patients told us they use pen-and-paper to collect their health data as they found it familiar and accessible. However, they described their handwritten data as ``\textit{really messy, \ldots sometimes [they] can’t even read [their] own handwriting or understand it}'' (P5).
To overcome this challenge, one of these patients tried using printed structured tables provided by clinics, but they said, ``\textit{explaining the table and numbers in it is not easy}'' (P9).  
One patient (P3) used several digital tools (i.e., glucose meters, smartphone apps, and smartwatches) to track their data. However, they found the tools not user-friendly and difficult to navigate, and the volume and complexity of the data were overwhelming. 
Four patients told us they do not actively collect any health-related data. One (P4) had initially used an app linked to a continuous glucose meter but found the detailed reports were, ``\textit{too much \ldots the data could probably be found within it, but it makes it more difficult to find it}'' (P4). As a result, they stopped collecting data altogether and relied on memory to observe changes in their health, noting that they sometimes forgot what they had observed. 

\section{Diary Study Results} \label{Results}
Our analysis revealed three themes (T1-T3).
This includes physical token selection and the rationale for choosing them (T1); evolving strategies for physically inputting health data (T2); and self-managing health through physical input data visualizations (T3). 
{
\renewcommand{\thesubsection}{T\arabic{subsection}}

\subsection{Patients’ selection of physical tokens and their rationale}
During the two-week study, patients collected various health-related data (i.e., diet, blood sugar, mood, weather). 
They represented these data using physical tokens that they found intuitive, memorable, and personally meaningful. 
Almost all patients arranged tokens in axis-based layouts to represent time by varying their position. Some manipulated the tokens’ length and size to convey duration or magnitude related to their data. Patients used tokens of different colors to reflect personal semantics, selected tokens that resembled real-world objects, and chose stickers for pictorial representations to make the visualizations easier to interpret and recall. We summarize the frequency of different mapping choices for the various types of health data collected across patients in \textbf{Figure~\ref{fig:encodings}}.

\begin{figure*}
    \centering
    \includegraphics[width=1.0\linewidth]{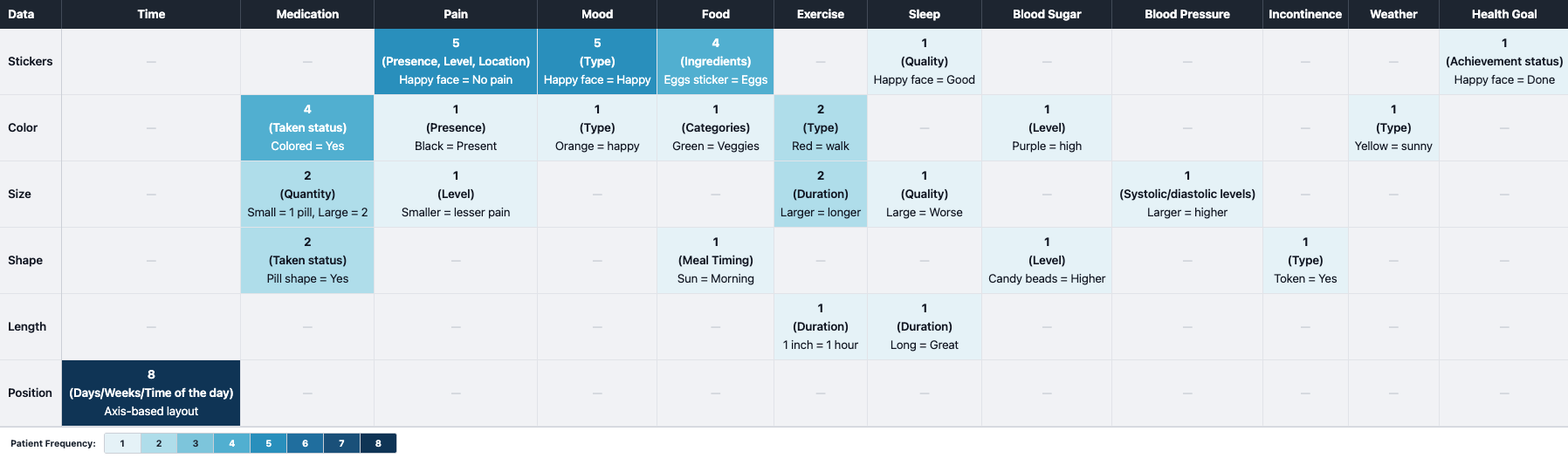}
    \caption{Frequency of different mapping choices for the various types of health data collected by patients.}
    \label{fig:encodings}
\end{figure*}

\textbf{T1.1: Arranging tokens in axis layouts to represent time: }
Most patients (8 out of 9) used axis-based layouts to represent time. They spatially arranged their visualizations, often creating a grid or plot where the x-axis or y-axis represented different data points over time. Patients mentioned they chose this representation because it mirrored familiar chart formats,  ``\textit{It’s basically a spreadsheet \ldots because I've been working on spreadsheets all my life}'' (P1).  
Four patients (P1, P5, P8, P9) used the x-axis to represent time (days of the week), while they used the y-axis to visualize health data points such as body pain, blood sugar levels, or food intake. 
One patient (P4) used the x-axis to represent days of the week and placed the time of the day on the y-axis, tracking activities such as walking and eating at specific times throughout the day. 
Two patients (P6, P7) placed the days of the week sequentially along the x-axis, and they did not assign any y-axis. They instead dedicated an entire column to a single day and placed health data points anywhere within that column without a specific order. 
P2 used the y-axis to represent time, with the x-axis for health data points such as pain level and amount of exercise. 

\textbf{T1.2: Manipulating length \& size of the tokens to represent duration, severity, \& count:}
Five out of nine participants manipulated the length and size of the tokens to represent differences in exercise and sleep duration, sleep quality, foot cramp intensity, and elevated blood pressure. They chose this approach because longer lengths and larger sizes naturally conveyed greater magnitude for them.
For example, P4 represented walking duration by cutting tape strips of different lengths, with each length corresponding to a specific number of steps, ``\textit{This strip is one inch equals 2000 steps \ldots 2.5 inches would be 5000 steps}'' (P4). They explained that this approach allowed them to scale the representation as needed, making it flexible and adaptable to their data. Some patients used tokens of different sizes to represent duration. For instance, they used differently sized woolen balls to indicate the quality of sleep, with larger balls representing bad sleep, ``\textit{This is how well I slept that night, here it [pointing to a large woolen ball] was particularly bad}'' (P1). 
Patients also used the size of tokens to represent severity and count.
For instance, they used woolen balls of varying sizes to track severity of blood pressure, with larger balls indicating critical spikes that required medical attention, ``\textit{So this one [pointing to a bigger ball on Sunday], my blood pressure was really high, and I had to go to the clinic and take my strong pill for that}'' (P5). P1 used  pins of different sizes to represent medication intake, explaining, ``\textit{If I take one, it's one pill [pointing to a smaller pin], and if I took two of them, then it would be a bigger pin.}''

\textbf{T1.3: Using colored tokens to associate with personal semantics: }
Five patients picked tokens of varying colors to represent their daily activities (e.g., food intake,  exercises) and emotions (e.g., calmness, happiness), as they associated specific colors with personal meanings. For example, P4 used green beads to represent consuming healthy food, ``\textit{And then the green is vegetables. Mm hmm. So, again, for me as a diabetic, vegetables are pretty well carb-free \ldots green healthy sort of thing}''. 
P3 picked tokens in different colors to present exercise types; they assigned red to indicate walking, as  ``\textit{Red to me is like friction. So, I'm walking on the ground and, you know, maybe that would mean something to me. Or it also means part of a heartbeat}'' and used blue to represent swimming.
While P8 picked colored tokens to represent emotions, ``\textit{For me, blue is a calm color. It gives off a sense of relaxation, which is why I thought it would be a good choice for these uneventful, calm days}'' and used orange to indicate days when they felt happy, ``\textit{I chose this particular color because orange always makes me think of joy and positivity. So, I used these small orange beads to indicate that I was happy and felt emotionally uplifted on that day}''.

\textbf{T1.4: Leveraging real-world metaphors for intuitive representations: }
Patients selected tokens that resembled the shape and texture of familiar objects from their daily lives, making intuitive representations that were personally meaningful and easy to remember.  
For example, P7 used cube-shaped beads for blood sugar medication because they resembled sugar cubes, ``\textit{And this one is for the pill I take to control my blood sugar. I picked this cube-shaped bead because it looks like a sugar cube.}'' Similarly, they used heart-shaped beads to represent medication for heart conditions, ``\textit{This one here, shaped like a heart, represents the pill I take for my heart condition. I chose this bead because it’s heart-shaped, which makes it easier to remember}'' (P7). P7 also selected small, round, glossy textured yellow beads to represent their cholesterol medications, as the beads reminded them of oil drops, which they associated with high cholesterol. Meanwhile, P5 used pink beads to represent their hypertension medications, ``\textit{So I chose these pink small beads because they look like the pills I take for my hypertension, my pills are small and in color pink}''. P6 used a glass bead to represent blood sugar levels due to its crystallized texture, ``\textit{So, this is nabat [a type of sugar commonly used in Middle Eastern and South Asian cuisine]. So, it looks like nabat, and nabat is sugar, so I chose this to show my blood sugar levels}'' (P6). 

\textbf{T1.5: Choosing pictorial stickers to directly map the data: }\newline
Patients used emoji stickers, body part stickers, and food stickers to represent their mood, pain levels, and meal data, as they thought these were easy to interpret and conveyed information without relying on written descriptions. Among the six patients who tracked pain, five used body part stickers to indicate both the location and the presence of pain. For example, P3 used bandaged body part stickers to represent pain in those areas, ``\textit{this is the brain\ldots it [stickers] was the best way to represent whether I had a headache. \ldots the Band-Aid on the brain is the headache.}'' P7 used a muscle sticker to indicate muscle cramps in their feet, explaining that it provided a clearer representation than writing it down, ``\textit{Using a muscle sticker would make it more obvious that I had issues with my muscles rather than writing it down.}'' 
Stickers were also commonly used for mood tracking, with five patients using emojis to visually express their emotional state. Four patients used food stickers to represent their food intake, ``\textit{The reason I chose it [food stickers] was because it was the best representation—it had meat, it had vegetables}'' (P3). Similarly, P7 used stickers for commonly eaten foods, stating, ``\textit{I mostly ate fruits and bread, so I used a strawberry sticker and a bread loaf sticker to show what I ate that day.}''

\subsection{Developing strategies for physically inputting health data}
Patients developed various strategies throughout the course of the diary study to input their health data visualizations using physical tokens. They experimented with ways to track their health data and decided what information to collect, how to represent it with tokens, and ways to make tracking an engaging part of their daily routines. They adapted these strategies to their unique needs, choosing token types and layouts that suited them best, and developed approaches to track health data on the go, staying flexible while creating and maintaining clear and meaningful visualizations.

\textbf{T2.1: Learning and experimenting with inputting health data: }
For many patients (4 out of 9), our study marked their first experience with regular, structured health tracking, having never consistently monitored their health data before. In the initial days, patients actively explored and tried to learn multiple aspects of personal health data tracking: determining which information to collect, selecting appropriate tokens, and developing routines for daily input. Some felt the newness was overwhelming, ``\textit{To cut them [tokens] up, to decide what to use \ldots It didn't work well for me}'' (P3). However, through tactile and playful interaction with tokens, curiosity and experimentation, patients quickly transformed unfamiliar tasks into motivating and enjoyable practices. Most patients (8 out of 9) found the hands-on process of collecting and representing data with physical tokens both engaging and enjoyable.
To decide what health data to track, some patients initially collected a wide range of information, anything they thought might be useful. Over time, they recognized that collecting irrelevant data sometimes obscured meaningful patterns, so they began prioritizing information most relevant for managing their conditions. As P5 explained, ``\textit{That [collecting more data] would be difficult. I would try to do my best to [collect meaningful data], but I would not record this one [data they thought not meaningful for their health] detail}''. \\
Patients also experimented with various ways to map their data using physical tokens, exploring shapes, sizes, and layout arrangements. As they tried new mappings each day, some felt it became challenging to remember what each token represented, particularly given age-related memory changes, ``\textit{I guess the biggest challenge is just remembering how it is [represented]}''(P1). To support recall and consistency, patients developed creative external aids, such as writing notes on sticky pads placed near their boards or making daily video recordings explaining their visualization mappings. 
As they became more comfortable with the process, patients tried to integrate data tracking into their daily routines. Some placed boards in frequently visited places in their homes so updates became effortless, ``\textit{I just kept it there [living room] that’s where it’s easy for me to see and update every day}'' (P4) and some paired data collection with moments like morning coffee or bedtime reflections. For several, the activity evolved into a reflective, mindful practice, incorporated into relaxation or meditation time, ``\textit{that [tracking data] is a mindfulness, that's a secret \ldots I felt more relaxed}''(P3). Through this process, patients developed personalized approaches that made tracking both consistent and rewarding, transforming the tracking data task into a motivating daily habit.

\textbf{T2.2: Evolving strategies to match unique needs:  }
When collecting and representing health data using physical tokens, patients demonstrated unique needs shaped by their age and the cognitive and physical demands of managing multiple chronic conditions. They developed creative strategies to match these needs when inputting health data with visualizations. 
Patients preferred large tokens, such as stickers, as these were easier to hold, manipulate, and interpret, given age-related changes in dexterity and vision. 
To optimize board space while accommodating these large tokens, they employed several adaptive strategies. 
Some sectioned the board, dedicating specific areas to particular data types to prevent overcrowding. 
Others reorganized their layouts by switching the x- and y-axes, using the longer side of the board to provide more space for token placement, ``\textit{I wanted those stickers since [they] were so big. But, yeah, I wouldn't have enough room, [so] I changed the layout}'' (P4). 
Patients wanted to use tokens that best reflected how they experienced their symptoms. They had a mental image of the location, intensity, or type of each symptom, and they wanted to see that same image represented on the board. 
For example, P5 tracked daily pain in specific body areas and wanted stickers that showed both the precise location and the level of pain: ``\textit{I could not find the sticker that exactly shows the areas and the level I want to point out}''. When patients could not find tokens that matched their intended meaning, they adapted creatively by using alternatives such as sticky notes or hand-drawn illustrations to represent their symptoms as closely as possible to how they felt them internally.

\textbf{T2.3: Supporting continuous health data tracking: }
Several patients (5 out of 9) were interested in carrying their physical boards with them in their daily lives, wanting to update their health data wherever they happened to be, whether at a picnic spot or while traveling. To do so, a few patients rolled up their boards along with the tokens, carried them in backpacks or bags, and unrolled them whenever they wanted to add new data. The ability to ``\textit{track on the go}'' encouraged patients to engage in healthy activities while staying mindful of their well-being. As P1 described, ``\textit{I could take it… I can update after walking.}''
However, as they began moving their boards more often, a few patients noticed that their tokens shifted due to repeated handling: ``\textit{I rolled it up. And then when I came back to do the Friday, [it] had fallen apart.}'' (P3) Patients used several strategies to preserve the structure of their visualizations. Some secured their tokens with small pins or tape before storing the boards, while others arranged tokens into distinct areas based on the days they collected data, ensuring that recording data for one day would not disturb other days’ entries. A few photographed their boards before putting them away, creating a quick visual reference to restore placement if anything moved.
These adaptations allowed patients to maintain both flexibility and continuity, enabling them to collect data wherever they went while keeping their visualizations intact.

\subsection{Impacts of input health data visualizations}


Inputting health data through visual representations served as a tangible reminder for self-care. \colortext{This process also facilitated both sensemaking and reflection. We differentiate between the two as sensemaking involves discovering patterns, connections, and relationships in one's health data, while reflection involves looking back at one's own behaviors and lifestyle choices in light of those patterns and considering what to maintain or change.}

\textbf{T3.1: Input visualizations, a tangible reminder for self-care: }
Patients often placed their physical boards for inputting health data with visual representations in highly visible, frequently used areas such as the dining room, living room, or near the kitchen, so that both they and their family members or caregivers would naturally encounter them throughout the day. The process of selecting tokens, inputting data, and reviewing the board in these everyday spaces served as a constant, passive reminder for patients to take care of their health.  
One mentioned the visualizations they created using tokens served as a reminder for them to take their medications by visually seeing the negative impact of missed doses, ``\textit{creating this board helped me to forget less to take my pills because I could see that [on] this day, I forgot it and it affected my next day readings}'' (P5). Another found the presence of the board a reminder to eat healthy and take their medications,  ``\textit{So, it was on the dining table, and I was seeing it every day and it kind of helped me to remember to take my pills or remember to be careful about what I am eating}'' (P6). 
Caregivers and family members also used the visualizations on the physical board to support patients in maintaining healthy self-care routines.
For example, one participant kept the board close to the kitchen so it was easily viewable by his wife while she prepared meals. She used it to check how many carbs, fruits, and vegetables he had consumed, and to monitor his daily steps, which helped them plan walks together to better manage his blood sugar levels. In another case, a patient’s wife, who was a nurse, regularly reviewed the board with him during their daily interactions to help identify patterns and connections between different data points. Together, they discussed strategies for improving his health and planning upcoming activities.

\textbf{T3.2: Sensemaking through input health data visualizations: } 
During our study, patients could find various relations and patterns in their health data. They mentioned having the data laid out on their board helped them understand their health better, ``\textit{Collecting my data in this way, where I can see all the ups and downs and understand what caused them .... 
When it’s just numbers, I don’t get the same sense of the changes, and I can’t see how much each number differs from the others when it’s just in the text. With this visual representation, I don’t even need my doctor to explain things to me because I can see for myself how eating cake or skipping my pills impacts my blood sugar}'' (P9). 
In another example, one noticed how sleep quality affected their pain levels, which then affects their ability to exercise and consequently their mood, ``\textit{If I have a terrible sleep, it affects my pain and affects my mood and my exercise}'' (P2). They also mentioned they found out how the weather impacted their well-being, ``\textit{If I had the good [weather,] I had a good sleep, no pain, my mood was better, and I exercised}'' (P2). One patient was able to see how excessive exercise affected their foot cramps (P1), and another saw that excessive exercise led to unintended low blood sugar when on increased insulin (P4).
One patient found out that their mood affects their blood pressure, ``\textit{This shows that in that day I was so angry (pointing to angry emoji) I had some conflicts with my family members and that made me angry. Which caused my blood pressure to go up}'' (P5).  

\textbf{T3.3: Reflecting on health-related behaviors via visualizations:}
Patients could reflect on their behaviours by reviewing their input visualizations. One patient reflected on how increased walking, eating healthier, and keeping on top of their medications led to lower blood sugar levels and thought about maintaining these habits to sustain the progress, ``\textit{I was really happy with that result [pointing to the happy emoji]. If I could stay in that range consistently, it would be amazing for my health, and I know my doctor would be pleased with me}'' (P9). 
Another patient reflected on their behaviours and particularly how yoga leads to having a clearer head, ``\textit{You can do all kinds of things without doing that yoga every day. But this tells me that \ldots if you do it, there might be a greater chance of having that clear head}'' (P3). 
Another patient mentioned that after seeing lapses in the activities track, they got motivated to resume exercising, ``\textit{I was a bad boy and didn't do much for exercising at home or I didn't go to the gym.  And I started going back to the gym and exercising at home in this last couple of weeks or so}'' (P1). 

}

\section{Discussion}
Here, we discuss the unique needs and preferences of older adults with MCC for health data input \colortext{and discuss the benefits and tradeoffs of using physical input visualizations for older adult patients with MCC}.
We also provide a set of design considerations to support the process of inputting health data with visualizations.

\subsection{Older adults’ needs and preferences for health data input}

In our study, we introduced older adults with MCC to input visualizations, where they input health data with visual representations using physical tokens.  
We did not provide any pre-set templates, as these could stifle creativity and limit patients’ ability to explore creative ways of mapping health data. 
Instead, we only offered a few example mappings for inputting data using physical tokens during the tutorial and in the cheat sheet.
When patients began inputting data, many were initially hesitant or uncertain about what data to collect, how to represent it, and what mapping choices would best suit their needs. 
Several started by replicating the mapping examples we provided during tutorials. This was understandable, given their limited experience with visualization and the physical and mental effort involved in inputting data with representations~\cite{Huron2014}.  
Over two weeks, patients got the chance to actively engage with the tokens, arranging, manipulating, and selecting them to represent their health data. 
This engagement helped them identify meaningful data to record and how to represent it, given their age and the demands of managing MCC. 
Some preferred larger tokens for visibility; others selected tokens resembling real-world objects from their daily routines for ease of interpretation. 
Gradually, patients moved beyond the example mappings to develop their own mapping strategies. By the end of the period, all patients had designed input visualizations of their health data that reflected their unique needs and individual mapping choices and found the process playful, enjoyable, and mindful.

The mapping choices patients made sometimes differed from conventional visualization practices. For instance, visualization research suggests that using multiple visual channels together to reinforce data by representing the same data in more than one way~\cite{Munzner2014}. However, in our study, when patients used multiple visual channels such as color and shape together, they did not use them intentionally for data reinforcement. Instead, they selected a single channel, such as shape, to represent data, while reserving other channels, such as color, exclusively for aesthetic purposes rather than data mapping. 
Similarly, although standard guidelines recommend identity channels for categorical data and magnitude channels for ordinal data, some patients used identity channels, such as 
color hue, 
to represent ordered attributes such as blood pressure level. 
Even though these approaches deviate from established practices~\cite{Munzner2014}, patients were able to input, interpret, and reflect on their data effectively. 
This raises important questions: \textit{To what extent do we need to educate older adults to develop their own input visualizations? How much emphasis should be placed on constructing the ``right'' visualizations if patients can interpret data and derive insights effectively using their own mapping strategies?} 
Considering the benefits of inputting data with visual representations using physical tokens for this population, we should explore ways to support older adults in approaches that are personally meaningful and meet their unique needs.

\subsection{Benefits and trade-offs of using physical input visualizations for health data tracking}
\colortext{Prior work has shown that older adults often perceive meticulous health tracking as burdensome ``illness work'' that adds to the everyday demands of managing their conditions~\cite{Ancker2015}. However, our findings demonstrate that the tangible, expressive, and personalizable nature of physical input visualizations can make this process enjoyable and even mindful, turning health data collection into a meaningful activity in its own right rather than an additional burden. While physical input visualizations offer these benefits, certain capabilities of digital tools, such as integrating sensor data or creating backups if damaged, would require additional materials, effort, and resources to achieve in physical form. In this research, however, we do not position physical input visualizations as a replacement for other health data tracking approaches. Rather, our findings suggest that physical input visualizations are particularly well-suited to supporting an exploratory phase of health data tracking, when individuals are learning about their conditions, identifying meaningful data, discovering triggers and patterns, and building awareness of the relationships between their behaviors and health outcomes.}

\colortext{Older adult patients with MCC are often asked by clinicians to track specific health indicators for a defined period to identify triggers or understand the impact of medications~\cite{Ploeg2019}. 
Physical input visualizations could support this type of focused exploration, where the engaging nature of the process can sustain motivation, and the visual accumulation of data supports sensemaking. 
In our study, participants initially collected a wide range of data, then gradually narrowed their focus to what health insights they were missing, what data was needed for that, and what was most relevant for managing their conditions. 
Once individuals have built this understanding and gathered insights about their health, for example, recognizing that certain foods spike their blood sugar, they may no longer need to track at the same intensity, as the knowledge gained through this period informs their ongoing self-care. Future work could build on this by exploring the appropriate duration and cadence of physical input visualization use for older adults with MCC,  and whether physical input visualizations could be reintroduced at key moments, such as when adjusting to new medications or experiencing changes in their conditions.}

\subsection{Design considerations to support the process of inputting health data visualizations for older adults with MCC}
Inspired by the results of our intro interview, diary study, and exit interview with older adult patients, we offer the following considerations for future research and for designers to support older adults in the process of inputting health data with visualizations. 

Older adults are prone to age-related declines in fine motor control and visual acuity over time~\cite{Liu2017}, which may impede their ability to input data.
We observed that most patients in our study used large tokens and tokens with high-contrast colors, as these were easier to grasp, position, and see. 
Some of our participants preferred using puffy balls to input their data as it was easy to grasp and hold for them.
This underscores the importance of considering the physical size and affordance of tokens as an important design parameter, favoring tokens that are easier to see, grasp, and arrange.
The soft foam pieces, puffy pompoms, and wooden blocks can be particularly suitable for patients with limited grip strength or hand tremors. These materials are lightweight and easier to hold without requiring precise finger control. 
\colortext{Tokens with built-in grips~\cite{Dominiak2024}, such as tokens with finger loops, could be offered to make placement and removal easier.} 
Textured tokens\colortext{~\cite{Lallemand2022}} can also be included to provide tactile cues, helping patients recognize and differentiate them without relying solely on vision. 
We invite future designers to explore the use of dynamically resizable physical tokens to help older adults track health data. 

\begin{dc}
{DC1:} Designers creating input visualizations for older adults to track health data should prioritize resizable and high-affordance tokens that are easy to view, grasp, and manipulate.
\end{dc}

Previous research has shown the benefits of reusing everyday objects as input tokens to meet individual needs and to enhance engagement, support self-reflection, and make data more intuitive~\cite{Karyda2021}.
Patients in our study also preferred tokens that resembled familiar everyday objects, such as their medications, as these made the representations intuitive, personally meaningful, and easy to recall. 
Thus, it is important to encourage patients to re-purpose everyday household items (such as expired pills, buttons, dried beans, cornflakes, fabric scraps, or bottle caps) as tokens empowering them to build a representation that feels familiar and personally meaningful. 
Research demonstrated the benefits of allowing people to make their own tokens or customize tokens to expressively represent their data based on their needs~\cite{Fabio2024}. 
Patients in our study wanted tokens that visually reflected how they mentally visualized their symptoms, for example, by mirroring their location in their body, and its perceived intensity.
When patients could not find tokens that matched their mental images, they adapted creatively, using sticky notes or hand-drawn illustrations, to convey their intended meanings. 
Moldable materials~\cite {Jansen2015} such as air-dry clay, play-dough, or wax can be provided to create personalized forms of tokens that represent how they experience their symptoms. 

\begin{dc}
 {DC2: }Future research should investigate ways to encourage re-purposing everyday objects and creating customizable tokens to support older adult patients with tracking and interpreting health data based on their needs.    
\end{dc}


For older adults, delays in recording can lead to incomplete or uncertain entries, as memory and attention may lapse if data is not entered immediately~\cite{Lim2019}. 
Patients in our study also expressed a desire to input their health data on the go, whether traveling, exercising, or spending time outdoors with family. 
Supporting on-the-go, unbounded~\cite{Bae2022makingData} health data input is therefore essential for maintaining continuity and accuracy in health tracking.
However, inputting data with visualizations while away from home can be challenging for older adults, as they must repeatedly pack and unpack boards and tokens, may struggle with their weight and bulk, and risk tokens shifting or losing their placement during movement.
To enhance portability of boards, physical boards could be designed as foldable or accordion-style, or constructed from flexible, rollable materials such as fabric, soft plastic, or rubber. 
These boards could include built-in compartments, Velcro, or snap-on attachments to secure tokens during transport. 
To improve token portability, options such as portable mini sticker printers for on-the-go token printing, reusable tokens to avoid carrying extra pieces~\cite{Morais2024}, or modular, stackable 2D pieces for compact storage~\cite{Dominiak2024} can be explored.
More broadly, future work could explore alternative representations and input materials to support portability, such as paper-based~\cite{Kiriphys2023} or other lightweight, foldable physicalizations that can expand for use and flatten for storage and transport. 
Hybrid approaches could also be explored, such as pairing a home-based physicalization with an accompanying mobile app that captures, stores, or synchronizes snapshots of physical token arrangements. Such designs could preserve the benefits of tangible, patient-driven representations while enabling older adults to record health data wherever they are.

\begin{dc}
{DC3:} Future studies should investigate lightweight, unbounded, and durable alternatives for boards and tokens that maintain the stability and integrity of input visualizations using physical tokens to support inputting health data on the go.  
\end{dc}

We started our study with a tutorial and gave participants a cheat sheet summarizing guides to assist them throughout the study.  
This resource served as a crucial starting point, particularly important given that many older adults with MCC had never regularly tracked their health data before and were unfamiliar with creating visual representations. 
While some initially tried to replicate the example mappings, everyone was able to create their own mapping choices within the two weeks, a notable achievement given their initial uncertainty~\cite{Fan2022UnderstandingHO}.
This highlights the importance of providing guidance, such as cheat sheets or tutorials, to help patients get started with health data input visualizations. \colortext{However, such guidance must be carefully balanced to inspire rather than constrain~\cite{Bae2022makingData}}.
For example, in our study, most participants (8/9) represented time by positioning their data along axes in a grid-like pattern. 
Such examples can serve as ideas for displaying temporal data, but rigidly prescribing fixed layouts (specific x and y axes, such as days versus time or days versus data types) should be avoided, as this could constrain creativity and limit personal agency~\cite{Huron2014constructiveVis}.
\colortext{To foster creativity and preserve personal agency, guidance materials should actively encourage exploration and experimentation~\cite{Edo2023}}, prompting patients to try different token types, and to adapt them according to their own needs and preferences. 
For example, cheat sheets could suggest alternative token materials, ways to represent different types of health data, or creative exercises for developing personalized mappings. They could also include reflective prompts such as \textit{``What changes would make this easier to use tomorrow?''} to encourage iteration and ongoing customization, supporting a process of continuous learning and adaptation.
\colortext{Guidance materials could also scaffold sensemaking and reflection. For instance, prompts such as \textit{What surprised you about your data today?''} could encourage patients to look for connections, questioning what might be connected, and considering what their data means for their daily routines.  }
\begin{dc}
{DC4:} Future research should develop guidance to help older adults create health data input visualizations while actively encouraging exploration and experimentation to foster creativity and preserve personal agency. \colortext{Such guidance should also prompt patients to engage with their emerging data, noticing patterns, questioning connections, and reflecting on what the data means for their self-care.}
\end{dc}


Patients in our study placed their input visualizations in frequently used spaces, such as the living or dining area, where these visualizations became a persistent presence in their environment. They valued this visibility as a constant reminder to input data;
however, as highlighted in previous research, it is important to preserve the privacy in visualizations~\cite{Sauve2024} since such visibility could raise privacy concerns: 
\textit{how can older adults be supported in preserving the privacy of their health data while still using physical tokens in familiar, frequently visited spaces? } 
Thus, we invite designers to rethink representations that enable older adults to selectively reveal or conceal health data whenever necessary. 
For example, reversible or double-sided boards could display detailed visualizations on one side and abstract or decorative patterns on the other, allowing patients to flip the board when guests are present. 
Modular covers, such as decorative fabric panels, sliding screens, or magnetic overlays, could quickly conceal detailed information. 
Privacy-aware tokens that reveal data only from certain angles or under specific lighting conditions could balance accessibility with discretion.
Researchers could also explore abstract representations resembling home decor~\cite{Kim2020, brombacher2023}, playful objects~\cite{Khot2014}, or jewellery~\cite{Stusak2014} to protect sensitive information. 
However, rather than providing entirely abstract representations, which may lead to oversimplification and misinterpretation~\cite{Liapis2019}, designers could create tokens that are rich with meaningful detail for patients while appearing as ordinary home items.  

\begin{dc}
{DC5:} Future designers should rethink representations that allow older adults to selectively reveal or conceal health data according to social context, personal comfort, and the level of detail they wish to share.
\end{dc}

\subsection{Limitations}
Our study had several limitations. 
We distributed our recruitment letters in several retirement and long-term care homes and in their newsletters to target diverse viewpoints. However, due to the challenges of recruiting patients~\cite{rajabiyazdi2021}, in particular older adults living with multiple health conditions, we could recruit 9 patients in a one-year time span. The heterogeneity of their conditions and backgrounds, while reflective of the MCC population, may also have introduced variability in their experiences and needs. 
As with any qualitative study, the analysis remains interpretative and may be subject to researcher bias. To mitigate this, three researchers independently analyzed and coded the transcripts before collaboratively discussing and resolving discrepancies. While this approach strengthens the reliability of our findings, some degree of subjectivity remains. However, similar to other qualitative studies, the rigour of our findings should be judged by their resonance and plausibility rather than generalizability~\cite{Schwarze2020}.

\section{Conclusion}
In this paper, we examined how older adults with MCC could use input visualization to track and meaningfully reflect on their health data. Through a comprehensive two-part study, we underscored the potential of physical input visualizations as a supportive tool for this population, emphasizing the importance of solutions that facilitate intuitive health tracking and deep reflection.
Our findings highlighted the diverse and unique needs of older adults living with MCC for tracking personal health data, illustrating how they adapt strategies and personalize input visualizations to align with their individual needs.
Moreover, our research illustrated how physical input visualizations act as tangible reminders for self-care and how the process of collecting data through these tokens facilitates sensemaking by helping older adults recognize patterns, discern relationships in their health data, and reflect on their behaviors.
Finally, based on these insights, we proposed actionable design considerations to support the process of
inputting health data with visualizations for older adults with MCC.
\clearpage
\section*{Acknowledgements}
We thank Dr. Wesley Willett and the iLab members for their valuable input during the 
brainstorming sessions. We are grateful to all study participants 
for their time and contributions. This research was supported by the Natural Sciences 
and Engineering Research Council of Canada (NSERC), under the funding number (RGPIN-2021-0422).

\bibliographystyle{eg-alpha-doi} 
\bibliography{Sections/ref}  
\end{document}